\documentclass[12pt,a4paper,titlepage]{article}
\usepackage[utf8]{inputenc}
\usepackage{amsmath}
\usepackage{amsfonts}
\usepackage{amssymb}
\usepackage{array}
\usepackage{caption}
\usepackage{enumerate}
\usepackage{graphicx}
\usepackage{multirow}
\usepackage{algorithm}
\usepackage{subfig}
\usepackage[noend]{algpseudocode}
\usepackage{float}

\usepackage{tabularx, booktabs}
\newcolumntype{C}{>{\centering\arraybackslash}p{2em}}

\usepackage[left=1in,right=1in,top=1in,bottom=1in]{geometry}

\usepackage{fancyhdr}
\usepackage{enumerate}

\makeatletter
\newcommand{\thickhline}{%
    \noalign {\ifnum 0=`}\fi \hrule height 1.5pt
    \futurelet \reserved@a \@xhline
}
\newcolumntype{"}{@{\hskip\tabcolsep\vrule width 1pt\hskip\tabcolsep}}
\makeatother

\title{Deduplication in a massive clinical note dataset}
\author{Sanjeev Shenoy$^1$, Tsung-Ting Kuo$^2$, Rodney Gabriel$^2$,\\
 Julian McAuley$^1$ and Chun-Nan Hsu$^2$\\\\
$^1$Department of Computer Science and Engineering and \\
$^2$Division of Biomedical Informatics\\
University of California, San Diego, La Jolla, CA.
}

\begin{document}
 \maketitle

\section*{Abstract}
Duplication, whether exact or partial, is a common issue in many datasets. 
In clinical notes data, duplication (and near duplication) can arise for many reasons, such as the pervasive use of templates, copy-pasting, or notes being generated by automated procedures.
A key challenge in removing such near duplicates is the size of such datasets; our own dataset 
consists of more than 10 million notes. 

To detect and correct such duplicates requires algorithms that both accurate and highly scalable.

We describe a solution
based on
Minhashing \cite{broder1997resemblance} with Locality Sensitive Hashing \cite{indyk1998approximate,gionis1999similarity}.
In this paper, 
we present the theory behind this method and present a database-inspired 
approach to make the method scalable. We also present a clustering technique using disjoint sets to produce dense clusters, which speeds up our algorithm.

\section{Introduction}
Clinical notes datasets are known to have 
significant
amounts of duplication. 
Detecting and `cleaning' data of such duplicates is necessary to perform correct and unbiased NLP analysis.
The large amount of notes in such datasets
(including our own)
rules out any na\"ive deduplication algorithm (for example one that is quadratic in the number of notes). Approximation algorithms aim to reduce the runtime
while bounding the accuracy to a certain threshold. 

Minhash was first used in the AltaVista search engine to detect duplicate web pages and eliminate them from search results \cite{broder2000min}. In the following sections, we describe the theory behind Minhash and how we have used it in deduplication of clinical notes.

\subsection{Goals}
Our primary requirements from an algorithm are speed and scalability. 
Under
these constraints, we 
seek an algorithm that is as accurate as possible in finding
near duplicate pairs. 
We will measure the accuracy of our algorithm in terms of the recall, i.e., the fraction of near duplicates that are correctly identified.

\subsection{Challenges}
The size of our dataset, and the large amount of exact and near duplicates in the dataset are the primary challenges of 
the problem. Existing open source software packages for document similarity clustering, to the best of our knowledge, 
are unable to scale datasets of the size we consider.

\subsection{Contributions}
Our contributions are a scalable deduplication algorithm using a database which can run on a single machine. We also maintain dense clusters using a modified version of disjoint sets during our process to make our algorithm 
faster.

\section{Document Similarity}
\subsection{Similarity Metric: Jaccard Similarity}
An important class of problems that Jaccard similarity addresses well is that of finding textually similar documents in a large corpus such as 
a collection of news articles \cite{strehl2000impact}. We use the Jaccard similarity measure to compute the similarity between two documents. For two sets, the value of similarity is the ratio of their intersection to their union.

An important point to note here is that Jaccard similarity operates on sets while a document can be considered to be a list, since words are ordered. We convert the document into a set by a shingling process which is described below. 

\subsection{Shingles and $n$-grams}

For some fixed $n$, 
\emph{shingling}
is the process of taking all $n$-length substrings from the document into a set. Punctuation and whitespace should not be ignored. Thus, each shingle contains exactly $n$ characters. 
An alternative
is to take contiguous sequences of $n$ words (instead of characters),
i.e., to use an $n$-gram based method.
We prefer 
the latter
since we can equate different forms of the same word while comparing documents.

We use the set of $n$-grams from the two documents to compute their Jaccard similarity. The reason we convert the documents to sets is that: 1) The probability that a document will have multiple instances of the same $n$-gram is low
in our corpus.
2) The order of the words is expressed in the $n$-gram. The order of the $n$-grams themselves will have low impact in determining near
duplication.

\section{Minhashing}
\subsection{Signatures}
Our goal in this section is to replace large sets by much smaller representations called ``signatures.'' The similarity-preserving property must hold; that is when the signatures of two sets are compared, their similarity must be close to the 
true similarity of the 
two documents.

\subsection{Minhash}

One of the techniques to produce a signature is minhashing \cite{broder1997resemblance}. Consider the sets of $n$-grams of all the documents in our dataset. We 
identify each $n$-gram with a
unique number. 
Next we
randomly permute this set of integers. For each document, its minhash value is equal to the minimum permuted value amongst all its $n$-grams.

\subsection{Minhashing and Jaccard Similarity}

The Jaccard similarity of two documents is equal to the probability that the minhash value of the first document is equal to the minhash in the second document \cite{broder1997resemblance}.
This property lies at the crux of this method. To compute Jaccard similarities we need to estimate this probability.

\subsection{Minhash Signatures}
The Minhash description above mentioned one random permutation for the set of $n$-grams. To calculate the probability of two documents having the same minhash value, we apply a fixed number $(M)$ of random permutations to produce $M$ minhash values. If $m$ is the number of trials for which the minhash values are equal, then we can approximate the probability as $m/M$.

Thus, for each document, we apply $M$ random permutations and for each permutation, we take the minimum value amongst its $n$-grams. As a result, we 
have reduced the representation of our documents to a set of $M$ numbers. We call this set a \emph{minhash signature}.

\subsection{Random Hash Functions}
Producing the $M$ random permutations on a large integer space can be 
computationally expensive. Hence, we need a much faster method of performing approximate random permutations. We use \emph{random hash functions} for this purpose. If the hash space is very large compared to the cardinality of the set of numbers being permuted, the probability that a pair of numbers will hash to the same number will be very low. Thus, each number in the set will be permuted to a unique number with a high probability.

Hashing is computationally much cheaper and it is easy to 
obtain
high accuracy by using a large space and hence, we use $M$ random hash functions to simulate the permutations.

\subsection{Deduplication with Minhash Signatures}
We have reduced each document to a much smaller representation with $M$ integers. However, finding the probabilities still requires us to consider each pair of documents and then comparing their minhash signatures to find the Jaccard similarity. The next step involves reducing the number of candidate pairs which we need to consider for calculating the Jaccard similarity.

For this, consider the signatures of two documents. The two documents will have a non-zero Jaccard similarity only when the signatures have at least one minhash value in common. Thus, we try to calculate such pairs and then for each such pair calculate the Jaccard similarity. We describe two possible methods for this computation.
\begin{enumerate}
    \item Create a list of documents for each minhash value produced. Each of these documents must have the minhash value in their signature.
    \begin{verbatim}
    Let L(v) be the (initially empty) list 
      which is hashed by minhash value v
    For each document d:
        Compute signature
        For each value v in signature:
            Add d to L(v)
    Let C be the (initially empty) list of candidate pairs
    For each value v:
        Add all pairs of documents in L(v) to C
    Compute Jaccard similarity for all pairs in C
    \end{verbatim}
    This method requires every signature value to be stored in memory. Thus the total memory requirement is 
    on the order of the number of documents times the
    number of hash functions; this requirement is too large and cannot be circumvented.
    
    \item Sorting the (minhash value, document) pairs and then computing candidate pairs
    \begin{verbatim}
    Let L be an empty list
    For each document d:
        Compute signature
        For each value v in signature:
            Add (v,d) to L
    Sort L
    Find the list of documents which have common values
    Compute candidate pairs for each such list
    Compute Jaccard similarity for all pairs in C
    \end{verbatim}
    This method achieves the same results as the first method. But there is no hard requirement on memory since the sorting can be done by external merge sort. However, external merge sort can be very slow. Another problem which affects both of these methods is that too many candidate pairs may be generated by this method since any document pair which has a similarity of $1/M$ has a $50\%$ chance of being discovered.
\end{enumerate}

\section{Locality Sensitive Hashing}
\subsection{Definitions}
\textbf{Hash Functions}: We have $M$ hash functions $h_1, h_2, ..., h_M$\\
\textbf{Signature Matrix}: Matrix with hash functions as the rows and the document signatures as the columns.

\subsection{Description}
\label{sec:5.2}
We now split the signature matrix into equally sized bands of rows. If we have $r$ rows in each band, then we will have $b=M/r$ bands. Now, we consider a pair of documents for similarity computation only if the two documents have the same signature in at least one band.

\begin{figure}[h!]
\centering
\includegraphics[scale=0.5]{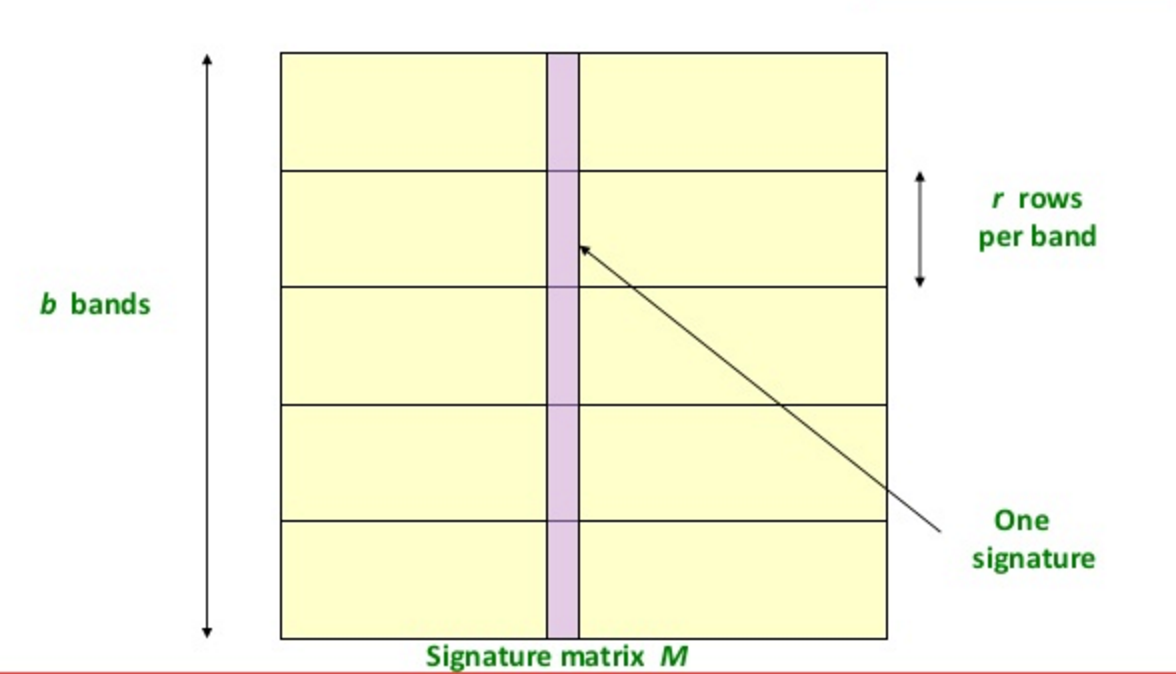}
\end{figure}

    \begin{verbatim}
    Compute the signature matrix
    Split the matrix into b bands each with r rows
    Let C be a (initially empty) list of candidate pairs
    For each band:
        Find all pairs of documents which have the same signature in the band
        Add these pairs to C
    
    Compute Jaccard similarity for all pairs in C
    \end{verbatim}
    
\subsection{Band Matrix}
\label{sec:5.3}
The signature matrix is too large to hold in memory. 
To circumvent this
we divide it into bands so that each band can be processed independently. However an entire band needs to fit inside memory
when
calculating the candidate pairs generated by a band. To minimize the amount of memory taken by the bands while keeping accuracy high, we hash the $r$ numbers in one band of a document to a single 64-bit number. Thus, one band can be represented by one integer per document. We call the matrix with these bands as rows 
a \textbf{band matrix}.

\subsection{Analysis}
Let $s$ be the Jaccard similarity between a given pair of documents. As we established earlier, the probability that
the minhash signatures for any two documents agree in any one particular row of the signature matrix is $s$. We can calculate the probability that these
documents (or rather their signatures) become a candidate pair as follows:
\begin{enumerate}
\item  The probability that the signatures agree in all rows of one particular band is $s^r$.
\item The probability that the signatures disagree in at least one row of a particular band is $1 - s^r$.
\item The probability that the signatures disagree in at least one row of each of the bands is $(1 - s^r)^b$.
\item The probability that the signatures agree in all the rows of at least one band, and therefore become a candidate pair, is $1 - (1 - s^r)^b$
\end{enumerate}
From the above analysis, it is expected that the probability of finding a candidate pair:
\begin{enumerate}
\item Increases with $s$, the jaccard similarity of the given pair
\item Increases with $b$, the number of bands
\item Decreases with $r$, the number of rows in each band
\end{enumerate}
Thus, the selection of $b$ and $r$ is closely tied to the threshold for similarity $t$. 
We describe this in further detail later.

\section{Storage Management}
To reduce our dependence on memory, we need to flush 
our processed values out of memory into the file system. Instead of handling the file system directly in our program, we make use of a database system to handle these aspects efficiently. For our purposes, we 
found Apache Cassandra 
to be a suitable choice.
Now we need a database design to efficiently store our processed values.
\subsection{Designs}
\label{sec:6.1}
\textbf{Design 1:}
For our first design, we have three columns:
\begin{enumerate}
    \item band\_id : An integer denoting the row in the band matrix
    \item document\_id : An integer denoting the column in the band matrix
    \item value: A Cassandra \emph{Bigint} for the 64 bit integer in the cell of the band matrix
\end{enumerate} 
The pair of band\_id and document\_id form the primary key of this databse table. In this design, for each value in the band matrix, we store it in a new row of the database table. The design is illustrated with an example. Consider the band matrix which has bands as the rows and documents as the columns (Table \ref{tab:band})

\begin{table}[htb]
\centering
  \begin{tabular}{| c | c | c | c | c |}
    \hline
     & $d_1$ & $d_2$ & $d_3$ & $d_4$ \\ \hline
    $B_1$ & 1 & 2 & 3 & 4 \\ \hline
    $B_2$ & 5 & 6 & 7 & 8\\ \hline
    $B_3$ & 9 & 10 & 11 & 12 \\ \hline
    $B_4$ & 13 & 14 & 15 & 16 \\
    \hline
  \end{tabular}
\caption{The band matrix \label{tab:band}}
\end{table}

This band matrix would stored as the following in the Design 1 table. 

\begin{table}[htb]
\centering
  \begin{tabular}{| c | c | c |}
    \hline
    band\_id & document\_id & value \\ \hline
    1 & 1 & 1 \\ \hline
    1 & 2 & 2 \\ \hline
    1 & 3 & 3 \\ \hline
    1 & 4 & 4 \\ \hline
    2 & 1 & 5 \\ \hline
    2 & 2 & 6 \\ \hline
    2 & 3 & 7 \\ \hline
    2 & 4 & 8 \\ \hline
    3 & 1 & 9 \\ \hline
    3 & 2 & 10 \\ \hline
    3 & 3 & 11 \\ \hline
    3 & 4 & 12 \\ \hline
    4 & 1 & 13 \\ \hline
    4 & 2 & 14 \\ \hline
    4 & 3 & 15 \\ \hline
    4 & 4 & 16 \\
    \hline
  \end{tabular}
\caption{Design 1 table}
\end{table}

\textbf{Design 2:}
Design 1 is simple but 
may result in too many rows for the database. 
Thus we
consider a design that will have 
fewer
rows. For this design, we first slice our band matrix by columns. Consider the example where we slice the band matrix such that each slice has two columns (Table \ref{tab:split}).

\begin{table}[h]
\centering
\begin{tabular}{C|C|C||C|C|C}
\cline{2-5}
& \multicolumn{2}{ c| }{band\_part 1} & \multicolumn{2}{ c| }{band\_part 2} \\ \cline{2-5}
& $d_1$ & $d_2$ & $d_3$ & $d_4$ \\ \cline{1-5}
\multicolumn{1}{ |C| }{$B_1$} & 1 & 2 & 3 & 4 &     \\ \cline{1-5}
\multicolumn{1}{ |C| }{$B_2$} & 5 & 6 & 7 & 8 &     \\ \cline{1-5}
\multicolumn{1}{ |C| }{$B_3$} & 9 & 10 & 11 & 12 &  \\ \cline{1-5}
\multicolumn{1}{ |C| }{$B_4$} & 13 & 14 & 15 & 16 &  \\ \cline{1-5}
\end{tabular}
\caption{The split band matrix \label{tab:split}}
\end{table}

Since we have sliced each band into parts, we call each slice a band\_part. Now instead of storing each cell of the band matrix as a separate row in the database table, we combine the columns to store each band\_part of the band matrix reducing the number of rows required. The details are listed below. We again have three columns for the database table:
\begin{enumerate}
    \item band\_id : An integer denoting the row in the band matrix
    \item band\_part\_id : An integer denoting the slice in the band matrix
    \item value: A Cassandra List of Bigints for the set of columns of each band
\end{enumerate} 
The pair of band\_part\_id and document\_id form the primary key of this database table. Again we illustrate this design with an example: The band matrix used above would be stored in our table as the following:

\begin{table}
\centering
  \begin{tabular}{| c | c | c |}
    \hline
    band\_part\_id & band\_id  & value \\ \hline
    1 & 1 & [1, 2] \\ \hline
    1 & 2 & [5, 6] \\ \hline
    1 & 3 & [9, 10] \\ \hline
    1 & 4 & [13, 14] \\ \hline
    2 & 1 & [3, 4] \\ \hline
    2 & 2 & [7, 8] \\ \hline
    2 & 3 & [11, 12] \\ \hline
    2 & 4 & [15, 16] \\
    \hline
  \end{tabular}
\caption{Design 2 table}
\end{table}

Thus, its clear that Design 2 combines few of the rows of design 1 into a single row to reduce the number of rows.

\subsection{Algorithms}
We need to add some more steps to our LSH algorithm 
in order to use the database correctly.
We describe the algorithms for both database designs.
\subsubsection{Design 1}
    \begin{verbatim}
    For each document d:
        Compute its band signature, i.e its column in the band matrix
        For each value in the band signature, insert into table
        (band_id, note_id, value) 
        
    Let C be the list of candidate pairs
    For each band_id:
        Retrieve the band with the id: select * from table where band_id = id
        Find all pairs of documents which have the same signature in the band
        Add these pairs to C
    
    Compute Jaccard similarity for all pairs in C
    \end{verbatim}
\subsubsection{Design 2}
    \begin{verbatim}
    Divide the list of documents into parts
    For each part P:
        Initialize each band to empty list B
        For each document d in P:
            Compute its band signature, i.e its column in the band matrix
            For each value in the band signature, append the value to its band B
        For each band B:
            Insert into table (band_id, band_part_id, B)
        
    Let C be the list of candidate pairs
    For each band_id:
        Retrieve the band parts with the id: select * from table where band_id = id
        Append all the lists retrieved above
        Find all pairs of documents which have the same signature in the band
        Add these pairs to C
    
    Compute Jaccard similarity for all pairs in C
    \end{verbatim}
    
\section{Clustering using Disjoint Sets}
A preliminary analysis of our original dataset revealed a 
significant
amount of duplication. Thus, there was a possibility that a large number of candidate pairs would be generated and evaluating them would slow down the algorithm considerably. We can tune $b$ and $r$ such that the threshold is high and low similarity pairs rarely show up as a candidate pair. However, we would also like to save some time by eliminating high similarity pairs while not losing the information that the pair was a high similarity pair.

Our solution for this is based on clustering.
The idea is that any pair of documents in a cluster is a high similarity pair and hence we needn't 
compute the Jaccard similarity exactly between all pairs in the cluster. As a result for large clusters, we are potentially removing a lot of candidate pairs. In the next section we describe a clustering method which guarantees that any pair of documents in a cluster has Jaccard similarity greater than a set threshold.
\subsubsection*{Disjoint Sets}
Disjoint Sets \cite{tarjan1979class} are a data structure used to represent sets. The operations supported are `find' and `union'. The sets are represented by trees. Both the operations of the disjoint set have a linear amortized time. This is because the depth of the tree is very small \cite{blum1986single}. We exploit this property of disjoint sets in providing a guarantee for Jaccard similarity between any pair of documents in a set.
\subsection{Triangle Inequality}
We know that the Jaccard distance follows the triangle inequality \cite{lipkus1999proof}. Let $j(A,B)$ be the Jaccard similarity between two documents $A$ and $B$. The Jaccard distance is then defined by $\delta(A,B) = 1 - j(A,B)$. Then, by the triangle inequality,
\begin{align*}
    \delta(A,B) + \delta(B,C) & \geq \delta(A,C)\\
    \Rightarrow 1-j(A,B)+1-j(B,C) & \geq 1-j(A,C)\\
    \Rightarrow j(A,C) & \geq j(A,B) + j(B,C) - 1
\end{align*}
Thus we have a lower bound on the Jaccard distance between A and C without actually calculating it.
\subsection{Extension to a Tree}
We can extend the inequality shown above to a tree. In a tree of documents, there is only one simple path between a given pair of documents. We can hence get a lower bound on the Jaccard similarity between any pair of documents in the tree if we know the exact Jaccard similarity between the pair of documents which form edges.
\subsection{Lower Bound Threshold Property}
Our aim is to maintain trees representing each set such that all pairs of documents in a tree would have a lower bound on the Jaccard similarity of at least the threshold. This threshold is called the \emph{tree threshold}.
\subsection{Extension of Disjoint Sets}
We extend the disjoint set data structure to include one more parameter: The minimum lower bound on the Jaccard similarity between the root and the leaves of the tree. The triangle inequality is being applied to all root to leaf paths and the one with least lower bound is saved. We now describe how we use this parameter in maintaining the Lower bound threshold property of the set. For this, notice that the set will change only during the union operation. Thus, no changes are needed for the find operation.
\begin{verbatim}
A node of the tree has the following attributes:
parent -> Parent in the tree
rank -> Height of the subtree defined by the node
min_score -> Minimum Jaccard similarity lower bound between the node 
and the leaves of the subtree defined by the node

Union(x,y)
    def union(x, y):
    xRoot = find(x)
    yRoot = find(y)

    if xRoot == yRoot:
         return

    sim = Exact Jaccard similarity between xRoot and yRoot
    leaf_to_leaf_min_score = xRoot.min_score + yRoot.min_score + sim - 2
    // Check if the lower bound threshold property still holds 
    if leaf_to_leaf_min_score < tree_threshold:
        return

    if xRoot.rank > yRoot.rank:
        yRoot.parent = xRoot
        xRoot.min_score = min(xRoot.min_score, yRoot.min_score - (1 - sim))
    elif xRoot.rank < yRoot.rank:
        xRoot.parent = yRoot
        yRoot.min_score = min(yRoot.min_score, xRoot.min_score - (1 - sim))
    else
        yRoot.parent = xRoot
        xRoot.rank = xRoot.rank + 1
        xRoot.min_score = min(xRoot.min_score, yProb - (1 - sim))
\end{verbatim}
Note that the union should be performed only when we can guarantee the Lower bound threshold property. This property is checked
by applying the inequality in the path between the leaves with least Jaccard similarity lower bound in the two trees. This is assuming that the edge joining the roots of trees has been added. \texttt{leaf\_to\_leaf\_min\_score} gets the lower bound for this path by applying the triangle inequality twice: once on the edge between the roots and once for concatenating the two root to leaf paths. It is clear that this is the least possible lower bound between the nodes of the two trees since we are picking the minimum from both trees.

Thus, we can 
perform the union only when \texttt{leaf\_to\_leaf\_min\_score} is at least the threshold so that we can guarantee that the property holds. Finally, we update the new \texttt{min\_score} of the root by applying the triangle inequality on the new edge that was added.
\subsection{Growing Disjoint Sets as Clusters}
\label{sec:7.5}
Our goal is to identify documents with at least one other near duplicate document that have Jaccard similarity scores as high as a given threshold in a data set. Consider that
all documents are connected as a graph with edges as their Jaccard similarity scores. This problem is equivalent to removing all edges with a score lower than the threshold. We therefore call this threshold the \emph{edge threshold}.

Now that we have described how to maintain sets and their useful properties, we show how it is used in our algorithm. Recall how the candidate pairs are generated. For each band, all pairs of documents which have the same signature in the band are considered as candidate pairs. We can then check if the pair actually has a Jaccard similarity score higher than the edge threshold. If so, we replace each document in the pair by the root document in the set in which the document belongs. The next step is to combine clusters using the \texttt{Union} of the sets. This is essentially the algorithm of the procedure \texttt{find\_candidate\_pairs}, which will be executed for each band. Clusters (disjoint sets) created in bands already processed will be inherited when a current band is processed and the procedure \texttt{Union} will merge two disjoint sets properly to ensure that the resulting set still maintains a mutual similarity at least as high as \emph{tree threshold}. Document pairs in each cluster will have the Jaccard similarity scores guaranteed to be higher than the given tree threshold. The guarantee serves our goal of identifying near duplicates. This algorithm also reduces the number of documents for which we need to actually evaluate their Jaccard similarity scores. 

\begin{verbatim}
find_candidate_pairs(Band B):
    Sort the band to identify sets of documents with the same signature value
    For each set of documents with the same signature value:
        For each document D in the set:
            replace D with D.find()    //This compresses the set
        For all document pairs (A,B) in the set:
            sim = Jaccard_Similarity(A,B)
            if sim > edge_threshold:
                Union(A,B)
\end{verbatim}

A pair with a Jaccard similarity score higher than the tree threshold may not always be in the same cluster and the algorithm will create different clusters depending on how bands and documents are ordered as its inputs. We may apply a minimum spanning tree algorithm to increase the chance for two highly similar documents to be merged into the same cluster. However, the time complexity will remain the same and the resulting clusters may still not be optimal if the triangle inequality bound is loose. We therefore implemented the clustering procedure as the algorithm given here.  

\section{Implementation}
We pin down some implementation details before we experiment with data.
\subsection{Dataset}
Our dataset consists of more than 10 million public domain clinical notes each of which contains a few hundred words. The size of the dataset is around 20 GB. The size of this 
dataset is the key challenge to deduplication. While we present our method for handling this dataset, the method is designed to handle a much larger dataset since scalability is one of the main requirements for the solution.
\subsection{Notes Preprocessing}
We have used Python for the implementation. First we split the documents into words and construct the $n$-grams.
The value of $n$ here is important since a high $n$ can result in similar documents 
being ignored
found while a lower $n$ can result in dissimilar documents being marked as similar. The value of $n$ that we have chosen is $8$. We have also 
pre-processed 
words by stemming them. Stemming reduces a word to its root form so that same words in different forms match.
\subsection{Hash functions}
 We use the MurmurHash \cite{appleby2008murmurhash} library which is a fast non-cryptographic hash function. We use the default random number generator to produce $M$ random seeds which are used to initialize the $M$ random hash functions.
 
\subsection{Hash Space}
To avoid collisions between different $n$-grams while hashing them, the hash space needs to be large. We have chosen our space to be as large as $2^{64}$ by using 64 bit numbers as our hash space. Even if we now have a billion documents each with some 1000 words in them, the total number of $n$-grams is $10^{12}$ which is much less than our hash space size.

\subsection{Various Implementations}
\subsubsection{Baseline}
For our baseline method, we simply compare all pairs of documents and calculate the Jaccard similarity. We apply a threshold on similarity to define 'similar' documents.
\subsubsection{Processing in Memory}
We implement the LSH algorithm in section~\ref{sec:5.2} entirely in memory. Instead of computing the signature matrix, we compute the band matrix. This compresses $r$ rows into a single row and hence is $r$ times more space efficient. This method requires us to have the entire band matrix in memory.
\subsubsection{Database designs}
We also implement both the database designs presented in section~\ref{sec:6.1}.

\section{Analysis of the Algorithms}
\label{sec:9}
Before we produce the results of running these algorithms on our dataset, we analyse the expected time and space complexity of the implementations discussed above.

1) \emph{Definitions}
\begin{itemize}
    \item $N$: Number of documents
    \item $w$: Maximum number of words in each document
    \item $M$: Number of hash functions
    \item $b$: Number of bands
    \item $r$: Number of rows in each band
    \item $p$: Number of band parts
    \item $d$: Number of documents in each part
\end{itemize}
Note that $M=b*r$ and $N=p*d$

2) \emph{Baseline}\\
\textbf{Time:} For each pair of documents, we compute Jaccard: $O(N^2w)$\\
\textbf{Space:} We don't need to store any document after it has been processed, hence the complexity is $O(w)$

3) \emph{Processing in Memory}\\
\textbf{Time:} For each document, we need to produce its signature by hashing each word $M$ times. This process takes $O(NwM)$ time. Producing the band matrix from the signature matrix takes $O(NM)$ time. Finally, we need to produce the set of candidate document pairs such that they have at least one common band value. For this, we have to parse each band once and collect all pairs which occur in each band. This step takes $O(Nb+P)$ where $P$ is the number of candidate pairs.

The value of $P$ depends completely
on the dataset.
Referring to Section~\ref{sec:5.3}, the probability that a given pair of documents $(i,j)$ is marked as a candidate pair is $1-(1-s_{ij}^r)^b$ where $s_{ij}$ is the Jaccard similarity between documents $(i,j)$. Thus, the expected number of candidate pairs that will be identified is $\sum_{i,j} 1-(1-s_{ij}^r)^b$. Since this is not a certain estimate, we rely on experimental data to give us more insights into this. This is explored in Section~\ref{sec:exp-clustering}\\
\textbf{Space:} We need to store the band matrix in memory. Thus, the complexity is $O(Nb)$

4) \emph{Database Design 1}\\
\textbf{Time:} For database designs, we need to consider the number of writes and the sizes of those writes to get an estimate into how much time it would take to run the program. The time complexity of the algorithm is the same as the previous one since we are just computing the same matrix and pushing it to 
disk. We now consider the database operations which add an extra overhead to the time.
\begin{enumerate}
    \item Number of writes: Each band matrix value is written as a row to the database : $O(Nb)$
    \item Size of each write: Each write consists of one row: (band\_id, document\_id and value). The total is $32+32+64=128$ bits.
    \item Number of reads: Each band is read once. The total number of reads is hence $b$.
    \item Size of each read: An entire band is copied from disk to memory from each read. The size is hence the size of each band which is $O(N)$
\end{enumerate}
\textbf{Space:} For each document, once we compute a band value, we immediately save it to disk. Hence this step doesn't require much memory. Getting the candidate pairs requires us to load the entire band matrix into memory. This is the bottleneck for memory for this algorithm. This step hence requires $O(N)$ memory.

5) \emph{Database Design 2}\\
\textbf{Time:} Again, we consider the database operations since the time complexity of all other operations is the same as the above two methods.
\begin{enumerate}
    \item Number of writes: We combine several columns of the band matrix and then write it. To be specific, we divide each row into p parts and write each part once to the database. Hence the total number of writes is: $O(pb)$
    \item Size of each write: Each write consists of one row: (band\_id, band\_part\_id and value). The value consists of $d$ 64 bit integers. Hence, the size of each write is $32 + 32 + d*64$ 
    \item Number of reads: Each band is read once. The total number of reads is hence $b$.
    \item Size of each read: An entire band is copied from disk to memory from each read. The size is hence the size of each band which is $O(N)$
\end{enumerate}
\textbf{Space:} For this method, we need to compute the band signatures of several documents and store them in memory before a write is made to flush these values to disk. Each part has to be entirely computed in memory. This step takes $O(db)$ memory. 
As with 
Design 1, to find candidate pairs, we need $O(N)$ memory. Hence, the total memory requirement is $O(db) + O(N)$.

6) \emph{Disjoint Sets}\\
\textbf{Time:} Disjoint sets have a constant amortized time for single operations. The number of times we do these operations is $O(N+P)$\\
\textbf{Space:} The space is linear in the number of documents: $O(N)$
\section{Experiments}
\subsection{Accuracy}
Before we perform experiments on measuring time and memory consumed by the different methods that we have described in the previous sections, we would like to fix the values of $b$ and $r$. As shown in section~\ref{sec:5.3}, the probability that a candidate pair would be found by the LSH algorithm is given by $1 - (1 - s^r)^b$. This function increases with $b$ and decreases with $r$. However we can't simply 
choose $b$ and $r$ based on this fact because these parameters affect other properties such as false positives, time and memory usage.

To measure the accuracy of the LSH methods, we need an
appropriate 
definition of `similarity'. We use Jaccard similarity with a threshold.
Whenever two documents have Jaccard similarity greater than a set threshold, we term them as similar.

\textbf{Test dataset:} We use a set of 521 clinical notes from i2b2/UTHealth 2014 challenge corpus~\cite{stubbs2015annotating,stubbs2015automated,stubbs2015heartdisease,stubbs2015identifying}. We also create near duplicates by changing 10\% of the words randomly in a given note. We create 10 such duplicates for a total of 531 notes.

We run the baseline method with different thresholds to define the list of near similar document pairs. Now the result is binary: for each pair, the LSH method has to label it as near-similar or not.

\textbf{False Positives:} Since we calculate Jaccard similarity for all the candidate pairs at the end of the algorithm, there would not be any false positives since the algorithm will discard those which have similarity less than the threshold. That is why we define false positives to be those which are in the candidate pairs but do not have similarity greater than the threshold. Thus, false positives are more like \emph{False Candidates} which cause our algorithm to run slower since more pairs have to be checked but otherwise doesn't affect the correctness of the algorithm.

Below, we test false positives and false negatives for different values of $b$ and $r$ and for different values of threshold.

\begin{figure}[htbp!]
    \centering
    \subfloat[False positives (candidates)]{{\includegraphics[scale=0.5]{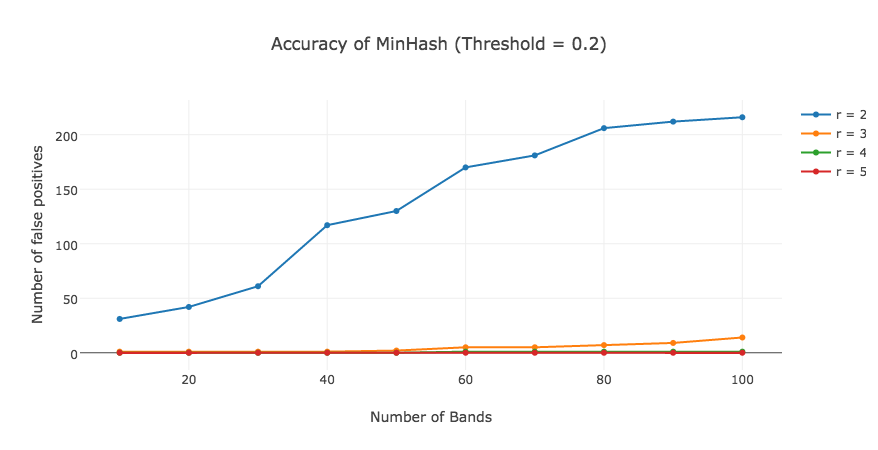} }}%
    \qquad
    \subfloat[False negatives]{{\includegraphics[scale=0.5]{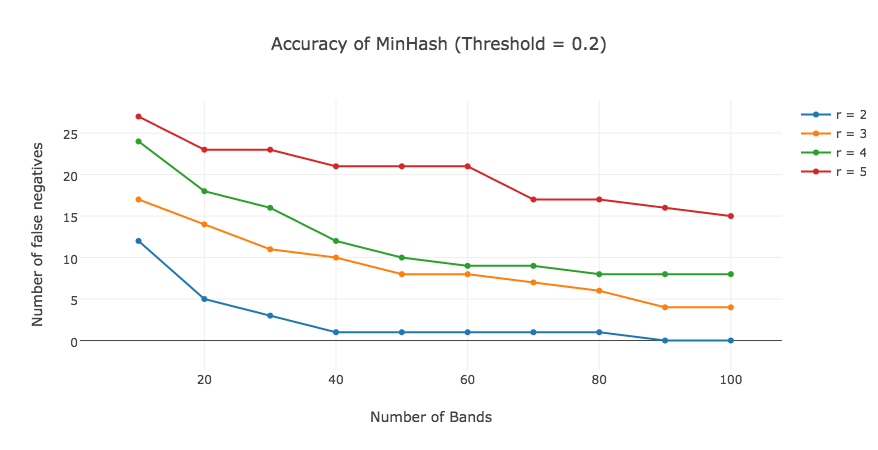} }}%
    \caption{Threshold = 0.2}%
    \label{fig:T02}%
\end{figure}
\begin{figure}[htbp!]
    \centering
    \subfloat[False positives (candidates)]{{\includegraphics[scale=0.5]{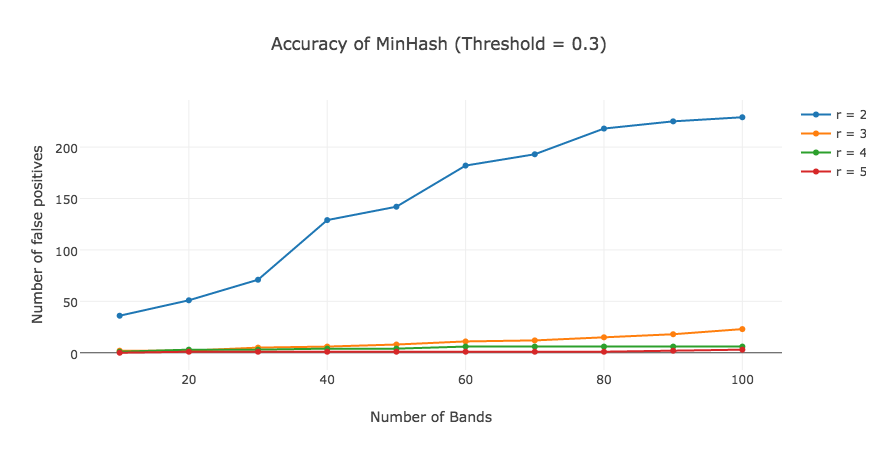} }}%
    \qquad
    \subfloat[False negatives]{{\includegraphics[scale=0.5]{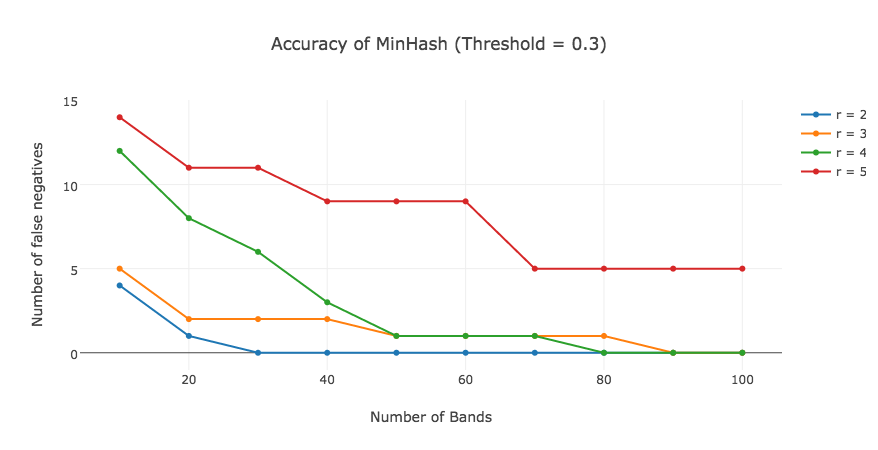} }}%
    \caption{Threshold = 0.3}%
    \label{fig:T03}%
\end{figure}
\begin{figure}[htbp!]
    \centering
    \subfloat[False positives (candidates)]{{\includegraphics[scale=0.5]{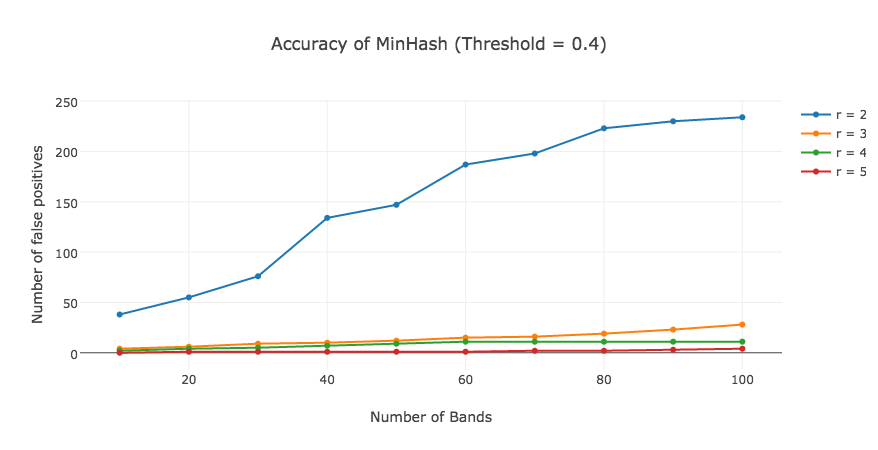} }}%
    \qquad
    \subfloat[False negatives]{{\includegraphics[scale=0.5]{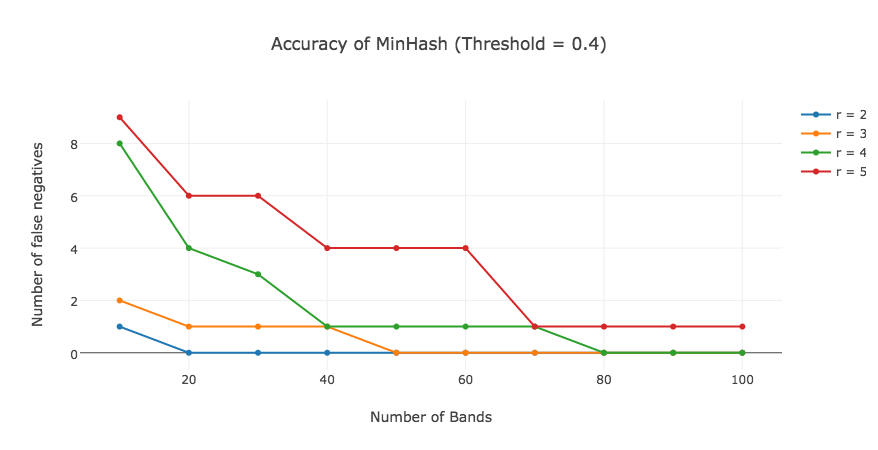} }}%
    \caption{Threshold = 0.4}%
    \label{fig:T04}%
\end{figure}

\begin{figure}[htbp!]
\centering
\includegraphics[scale=0.5]{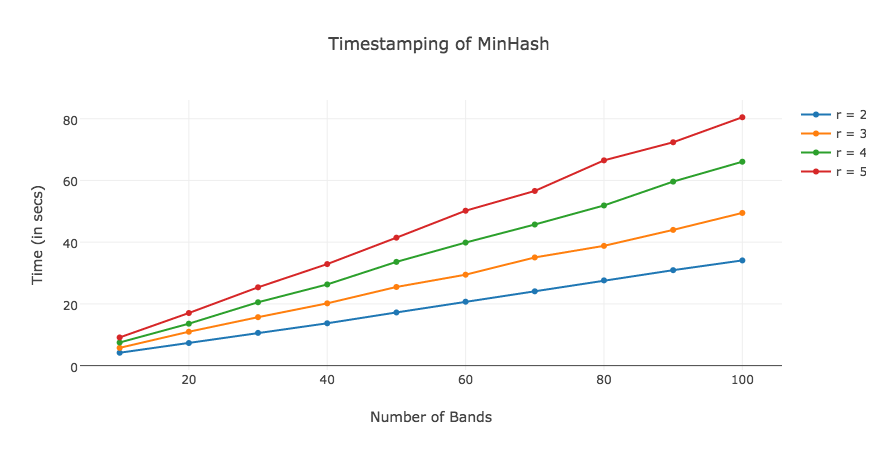}
\caption{Time taken}
\label{fig:loopgraph}
\end{figure}

\pagebreak The expected trends are clear:
\begin{itemize}
    \item False positives increase with threshold and false negatives decrease with threshold. This is because a larger threshold implies fewer pairs of documents are now considered similar. 
    \item An increase in the 
    number of bands which is also an increase in the hash functions increases the probability for a candidate pair to be found. Hence, false positives increase and false negatives decrease.
    \item An increase in the number of rows decreases the probability and hence an opposite effect is observed
\end{itemize}
Here we also show the time required to run the in-memory implementation of LSH on the given dataset for all the parameters (Figure \ref{fig:loopgraph}).
The results above also show us that the number of false positives in the worst case is around 250. The number of documents is 521. Thus, the number of candidate pairs being found is an order of magnitude smaller than the total number of pairs. Also for our purposes, avoiding false negatives is important. This allows us to select a configuration that completely avoids false negatives while allowing some false positives. Hence, for the remainder of the experiments, we have chosen $r=2$ and $b=50$.
\subsection{Time}
Here we compare the time taken by the in-memory implementation versus the two database designs. Section~\ref{sec:9} shows us that the algorithm is linear with respect to the number of documents, the average number of words and the number of bands present. Figure \ref{fig:loopgraph} shows us the linearity with respect to the number of bands. We also considered the number of database reads and writes and also the sizes of each of those reads and writes. We now benchmark our implementations on different numbers of notes to see the dependence on the database and how much the operations costs.

The only parameter that we have not already set is $p$ which is the number of parts which is required for the Design 2 implementation. Here we choose $p = 10$.

\begin{figure}[htbp!]
\centering
\includegraphics[scale=0.5]{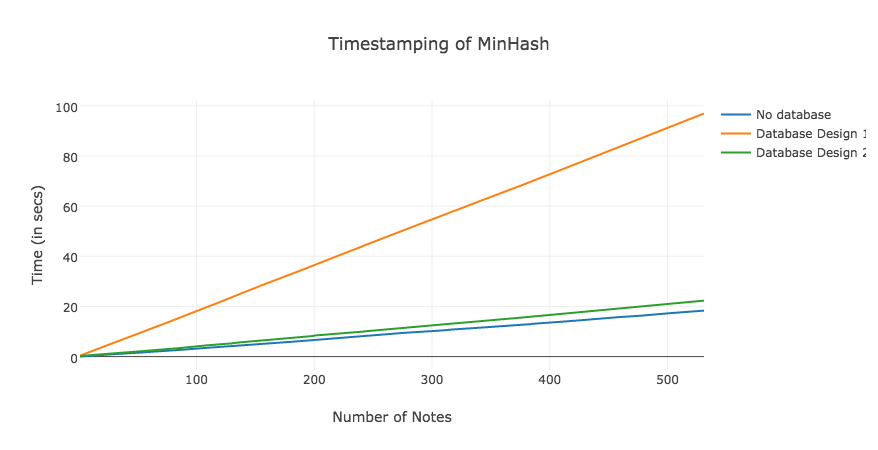}
\caption{Time taken}
\label{fig:time1}
\end{figure}
\begin{figure}[htbp!]
\centering
\includegraphics[scale=0.5]{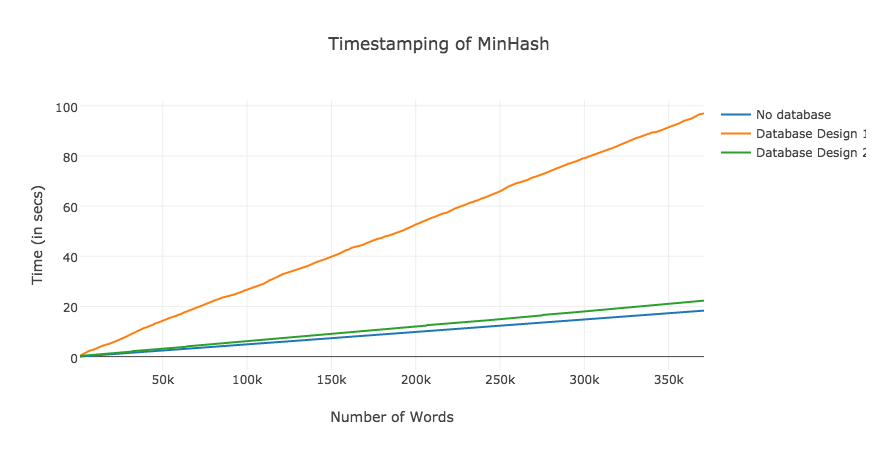}
\caption{Time taken}
\label{fig:time2}
\end{figure}
Figure \ref{fig:time1} shows us that because of the reduction in database writes, Database Design 2 performs considerably better than Design 1 while executing at a speed close to the in-memory implementation. The graph also shows linearity with respect to the number of notes. Figure \ref{fig:time2} suggests that 
the running time depends more on
the number of notes compared to the number of words.

For Design 2, we make writes for every 50 documents processed. Hence there are spikes in the time graph after every 50 documents. The spikes would grow much larger when there are more 
documents.
\begin{figure}[htbp!]
\centering
\includegraphics[scale=0.5]{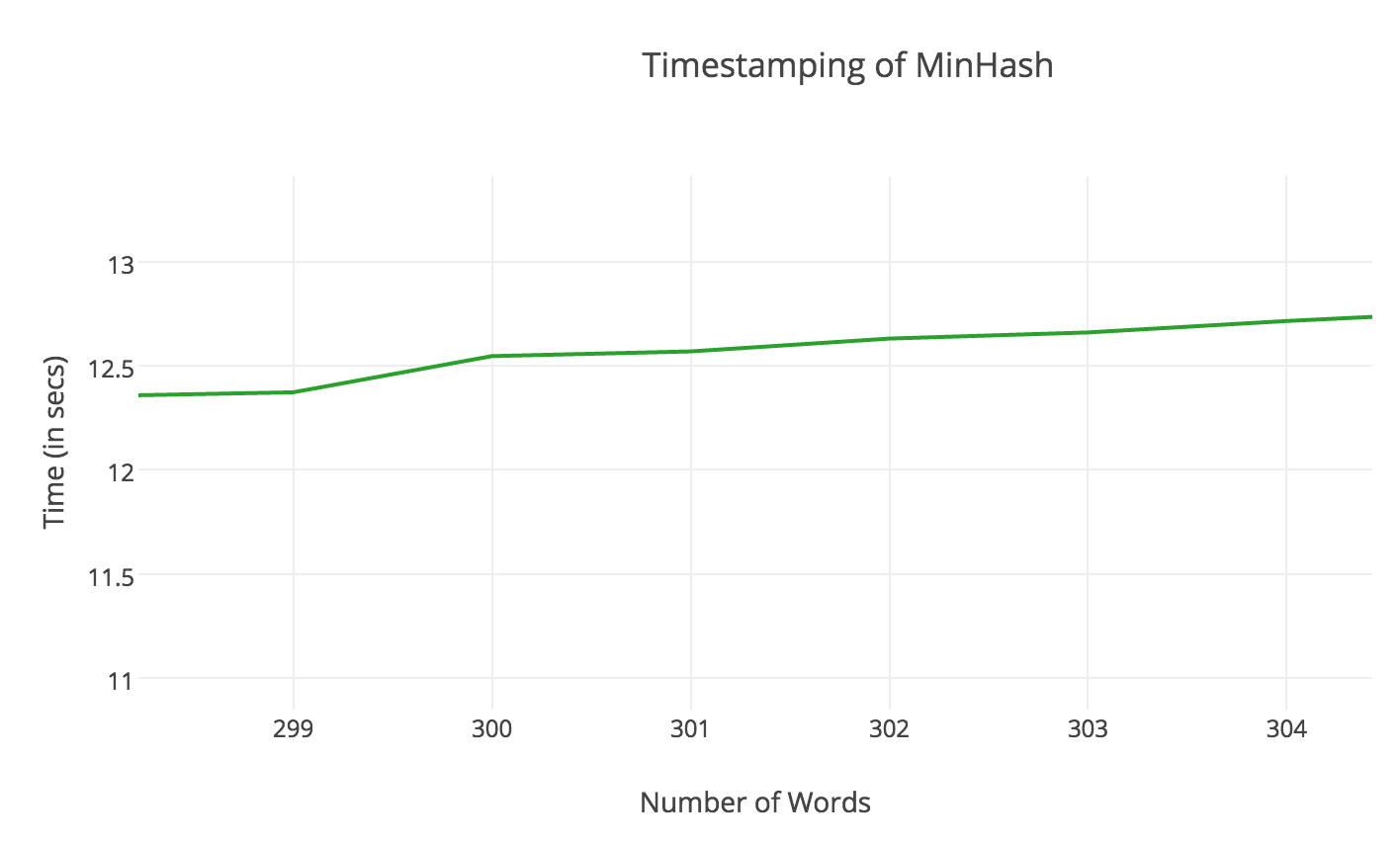}
\caption{Spike in graph at 300 for Design 2}
\label{fig:time3}
\end{figure}

\subsubsection{Memory}
\begin{figure}[htbp!]
\centering
\includegraphics[scale=0.3]{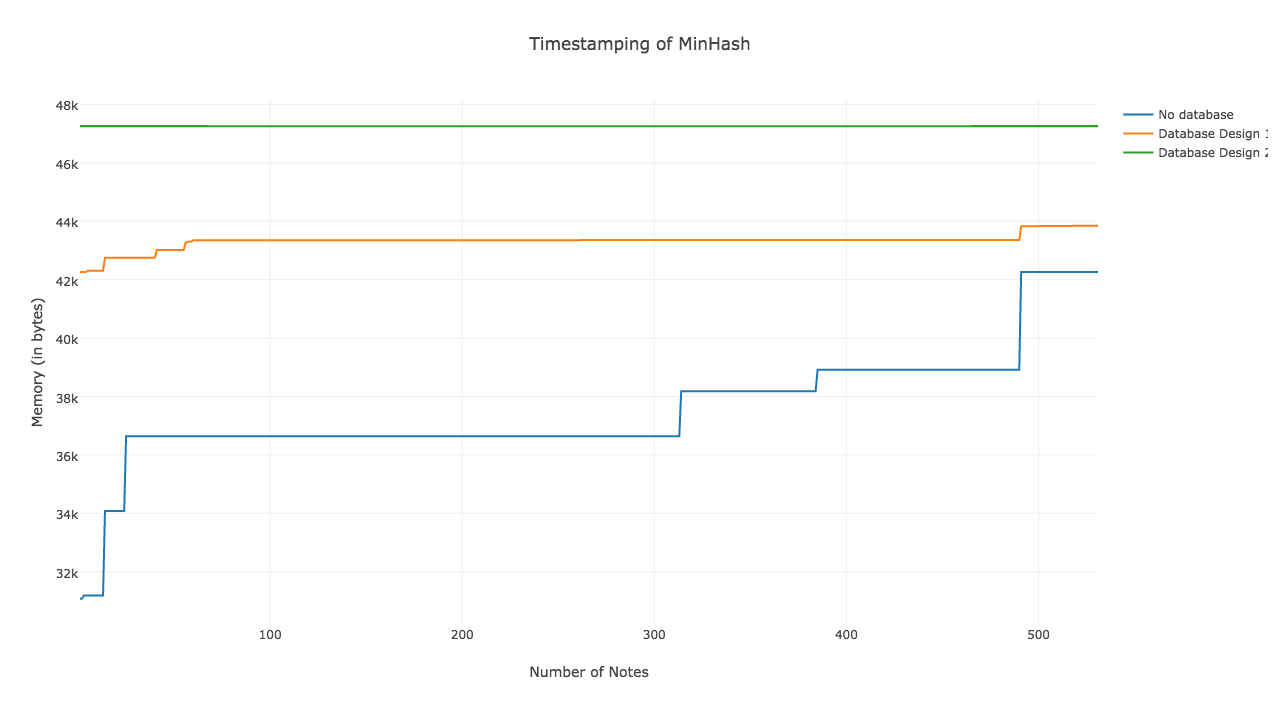}
\caption{Memory}
\label{fig:mem1}
\end{figure}

We saw that Design 2 is substantially more time efficient. However, it needs significantly more memory as we discussed in Section~\ref{sec:9}. We measure the memory allocation for our Python script for different implementations on the given dataset (Figure \ref{fig:mem1}).
The experiment shows us that the memory usage of the in-memory implementation increases as the number of documents increases. The reason that we don't get consistent values
is due to the memory allocation carried out by the interpreter which allocates more memory than required. The process is repeated when memory runs out.

\section{Experiments on Clustering}
\label{sec:exp-clustering}
We perform experiments on a sample dataset to test the parameters of the algorithm as well as analyse the kind of clusters that are formed.

\textbf{Sample Dataset:} We use the same dataset as before but this time, we inject a large number of near duplicates by artificially creating them. For this, we pick a random document and randomly change some of 
its words.
We repeat this procedure 500 times. The number of words changed varies from 0\% to 20\% so that we get a range of Jaccard similarities. With this, we end up with a dataset of 1021 documents.

\textbf{Tree Threshold:} The tree threshold should be a value such that all documents in a cluster have some common $n$-grams such that some useful information can be derived. A low threshold will produce larger clusters which might make information extraction more difficult while a large threshold will result in very small clusters. For our purposes, we have used a tree threshold of 40\%.

\textbf{Edge threshold:} The edge threshold is again a key factor 
that determines the resulting clusters.
We now perform 
experiments to help us select a good value of this
threshold.
\subsection{Varying edge threshold}
We run the algorithm defined in Section~\ref{sec:7.5} for different values of the
edge threshold on the dataset described above. We also run the original algorithm without using disjoint sets for comparison.

\textbf{Number of Jaccard computations saved:} Since one of our main goals is to make the algorithm faster by avoiding computing excessive pairs, we first measure how many pairs for which we avoided computing Jaccard similarity, for different edge thresholds
(Table \ref{tab:eliminatePairs}).
The number of pairs generated without using disjoint sets is 6388.
\begin{table}[htb]
\centering
\begin{tabular}{|c|c|} 
 \hline
 \multicolumn{2}{| c |}{\textbf{Number of pairs excluded}}\\
 \hline
 Edge Threshold (\%) & Pairs\\ \hline
 60  & 3398 \\ \hline
 65  & 3440 \\ \hline
 70  & 3428 \\ \hline
 75  & 3409 \\ \hline
 80  & 2696 \\ \hline
 85  & 2409 \\ \hline
 90  & 1837 \\ \hline
 95  & 1707 \\
 \hline
\end{tabular}
\caption{Number of pairs excluded from considering with different edge thresholds.}
\label{tab:eliminatePairs}
\end{table}

\begin{figure}[htbp!]
\centering
\includegraphics[scale=0.4]{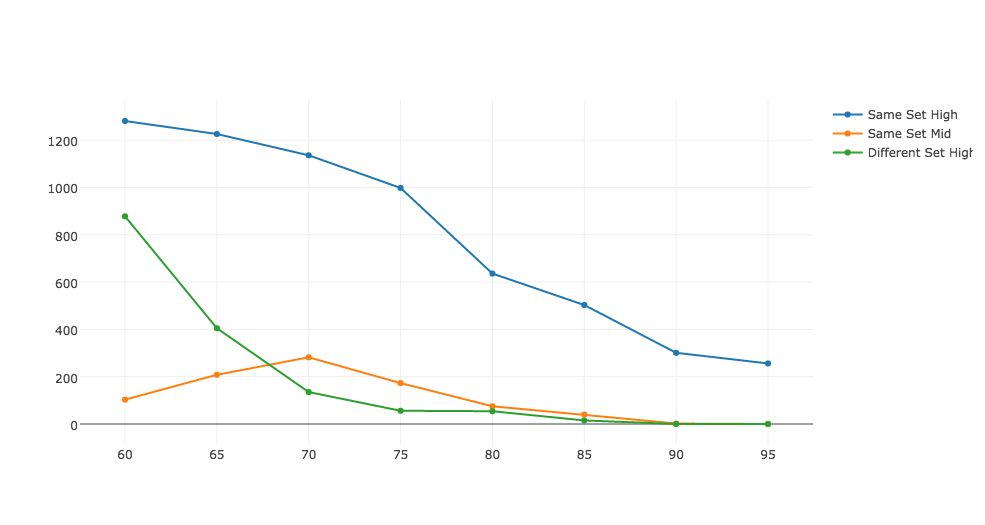}
\caption{Candidate Pair distribution (y-axis) as a function of varying edge threshold (x-axis).}
\label{fig:et1}
\end{figure}

We divide the pairs into three more categories according to their similarity and how they are clustered:
\begin{enumerate}
\item Number of high similarity pairs (Jaccard similarity $>$ edge threshold) belonging to the same cluster.
\item Number of high similarity pairs (Jaccard similarity $>$ edge threshold) belonging to different clusters.
\item Number of mid-similarity pairs (tree threshold $<$ Jaccard similarity $<$ edge threshold) belonging to the same cluster. 
\end{enumerate}
Figure~\ref{fig:et1} shows these numbers as functions of varying edge thresholds. The figure shows that if the edge threshold is low, it is expected that a large number of highly similar documents will be merged into the same clusters (category 1, blue curve) but it is also more likely for the lower bounds of two clusters to exceed the tree thresholds resulting in over-partitioning of a large number of highly similar documents in different clusters (category 2, green curve). On the other hand, when the edge threshold is as high as 80\%, nearly all highly similar documents are correctly merged into the same clusters, which is ideal for identifying duplicates. The number of document pairs of medium similarity (category 3, orange curve) grows initially and reaches its peak when the edge threshold is 70\% before it starts to decline and flatten out as the edge threshold increases. An optimal setting of the tree threshold for a low edge threshold to minimize the number of pairs of highly similar documents but in different clusters may depend on the data set. If necessary, we may apply a second round of clustering to merge clusters with highly similar documents.  

\textbf{Modularity:} Modularity~\cite{newman2006modularity} is 
a measure of graph structure that is commonly used to evaluate community detection algorithms.
It takes a value between -1 and 1 that measures the density of links inside communities compared to links between communities. For a weighted graph, modularity is defined as:
$Q = \frac{1}{2m}\Sigma_{ij}\bigg[A_{ij} - \frac{k_i k_j}{2m}\bigg]\delta (c_i,c_j)$. Where $A_{ij}$ represents the edge weight between nodes $i$ and $j$. $k_i$ and $k_j$ are the sum of the weights of the edges attached to nodes $i$ and $j$ respectively. $m$ is half the sum of all edge weights in the graph. $c_i$ and $c_j$ are the communities of the nodes, and $\delta$ is $1$ when $c_i=c_j$ (or $0$ otherwise). We calculate the modularity for each of the clusters generated, using Jaccard similarity scores as the weights between nodes (documents). We also show the number of clusters with at least two documents (Table~\ref{tab:clusterAnalysis}).
\begin{table}[htb]
\centering
\begin{tabular}{|c|c|c|} 
 \hline
 \multicolumn{3}{| c |}{\textbf{Cluster Analysis}}\\
 \hline
 Edge Threshold (\%) & Modularity & Number of clusters\\ \hline
 60  & 0.156 & 55 \\ \hline
 65  & 0.162 & 49 \\ \hline
 70  & 0.161 & 46 \\ \hline
 75  & 0.159 & 45 \\ \hline
 80  & 0.090 & 45 \\ \hline
 85  & 0.071 & 45 \\ \hline
 90  & 0.042 & 45 \\ \hline
 95  & 0.036 & 45 \\
 \hline
\end{tabular}
\caption{Modularity and number of clusters generated with different edge thresholds.}
\label{tab:clusterAnalysis}
\end{table}
\subsection{Comparison with Louvain method}
Many of the community detection algorithms try to optimize modularity. The Louvain Method \cite{blondel2008fast} for community detection is a method to extract communities from large networks. The method is a greedy optimization method that runs in time $O(N \log N)$.

We ran the Louvain method on the graph constructed using the Jaccard similarity scores returned by the minhashing algorithm without disjoint sets.

To compare with the Louvain method, we choose an edge threshold of 75. 
This setting leads to a high number of same-set high similarity pairs but a small number of different-set high similarity pairs.

For the Louvain method, the number of same-set high similarity pairs is slightly less than the disjoint set method. However, the number of same-set mid and low ($< 40$\%) similarity pairs is very high. For the disjoint set method, the mid similarity pairs are 173 and low similarity pairs are zero because of the 
set 
lower bound threshold property.

The Louvain performs well when we consider high similarity pairs belonging to different sets. No such pairs were found in the clustering done by Louvain. Our method produces 56 pairs. Finally, the modularity score is much higher for the Louvain method.

Based on the sets generated, our method produces 
denser clusters than the Louvain clusters. This makes it easier to analyse the cluster to determine common elements. Our method also preserves the edges between clusters and hence it is possible for any other algorithm to combine clusters for further analysis. This way, the number of high similarity pairs between two different sets can be reduced. Finally, our method has a large number 
of clusters
of size one. This is because such documents do not have significant similarity to other documents. The Louvain method puts such documents into larger clusters which can be meaningless for further analysis.

The main advantage of our algorithm is in the number of Jaccard similarity pair computations that we eliminate. For the edge threshold of 75, the number is 3409 which is about 53\% of the total pairs. This enables us to get a speedup of more than 2 times. In the clusters generated, more than 80\% of the document pairs are high similarity pairs. This will make it easy to derive conclusions about why the cluster has similar documents.

\begin{table}[h!]
\centering
\begin{tabular}{|c|c|c|} 
 \hline
 & & \textbf{Disjoint Set Clustering}\\ 
  & \textbf{Louvain Method} & (edge threshold$=75\%$) \\ \hline
 Same Set High  & 954 & $>$ 1000 \\ \hline
 Same Set Mid  & 3741 & 173 \\ \hline
 Same Set Low  & 1693 & 0 \\ \hline
 Diff Set High  & 0 & 56 \\ \hline
 Modularity  & 0.490 & 0.159 \\ \hline 
 Number of clusters & 57 & 45 \\
 \hline
\end{tabular}
\caption{Experimental comparison with the Louvain Method.}
\label{tab:louvain}
\end{table}

\section{Scalability analysis}
\subsection{Memory Analysis}
Finally, we conclude by computing the limit on the amount of notes that ca be processed by the different implementations which are constrained by memory. 
We assume a commodity machine with 4GB of main memory.
\subsubsection{Processing in Memory}
The band matrix needs to be stored in memory. Each value in the matrix is 64 bits = 8 bytes. The size of the matrix is $b\times N$ where $b$ is the number of bands and $N$ is the number of documents. Since we have chosen $b$ to be 50 in our implementation, the size of the band matrix is hence $8*50N = 400N$ bytes. Thus, the theoretical bound on the number of documents that can be processed with the given parameters is 4GB/400 bytes which is around 10 million.
\subsubsection{Database Design 1}
This implementation only requires each band to be in memory at a time. The size of each band is $N*64$ bit integers which is $8N$ bytes. Thus, Database Design 1 can handle around 500 million notes.
\subsubsection{Database Design 2}
This implementation requires a part to be in memory at a time. The memory requirement is $d\times b$ 64 bit integers where $d$ is the number of documents in each part. If we have 10 parts like in the experiments, the memory requirement is $b\times N/10$. Thus, the memory requirement is 10 times less than the in-memory implementation. 
Hence we can process at most 100 million notes with this implementation.

\section{Processing Real Clinical Note Data}
We ran the final Minhash algorithm with the disjoint sets as described in previous sections to process the entire corpus of the clinical notes from the medical center of UC San Diego, from 2012 to 2015. Some of the unusual things encountered were non-consecutive note ids and notes with fewer than 4 words. Creating signatures for the notes and writing them to the database took around 75 hours to complete. Reading the bands and finding candidate pairs and evaluating Jaccard similarity for the pairs with the disjoint sets took another 24 hours. Since the second step is not too long, this can be repeated for different edge thresholds to get other interesting results.

The total number of generated pairs after running the algorithm were around 341 million. The algorithm also generated 6,160 clusters of exactly similar notes. The largest of these had 2,540 notes. The number of clusters generated by our disjoint set algorithm with an edge threshold of 75\% was 13,233. The largest cluster had 9,849 notes.

\section{Conclusion}
We have presented implementations of LSH which vary in time and memory consumed but produce the same results. Our requirement was for an approximate but accurate method which can process a large amount of clinical notes while being as fast as possible. Our experiments show that LSH with Database Design 2 serves all our purposes while being slower than the in memory implementation by only a small percentage. We also described a method that clusters the notes as the pairs are generated to reduce the total number of pairs generated. The method produces high density clusters which can be useful for further analysis.

\textbf{Future work:} 
We plan to
analyse clusters and discover reasons 
that cause high-density clusters to occur.
Next, we seek to produce `clean' datasets
by removing redundant notes so that there is an unbiased dataset which can be used with different text based machine learning algorithms.

\bibliographystyle{unsrt}
\bibliography{Sanjeev}
\end{document}